\documentclass[prb,aps,groupedaddress,floatfix,superscriptaddress,twocolumn]{revtex4-2}
\usepackage[utf8]{inputenc}
\usepackage[T1]{fontenc}
\usepackage{graphicx}
\usepackage{amssymb}
\usepackage{amsmath}
\usepackage[dvipsnames]{xcolor}
\usepackage[colorlinks=true,linkcolor=blue]{hyperref}

\newcommand{\figref}[1]{Fig.~\ref{fig:#1}}
\newcommand{\equref}[1]{Eq.~(\ref{eq:#1})}

\newcommand{\e}{\rm e}

\newcommand{\mP}{\mathcal{P}}

\newcommand{\up}{{\uparrow}}
\newcommand{\down}{{\downarrow}}
\newcommand{\ev}[1]{\left\langle#1\right\rangle}

\newcommand{\sz}[1]{\hat{S}^z_{#1}}  % sz operator
\renewcommand{\sp}[1]{\hat{S}^+_{#1}}  % s+ operator
\newcommand{\sm}[1]{\hat{S}^-_{#1}}  % s- operator
\renewcommand{\i}{\rm i}
\newcommand{\ani}[1]{\hat{c}_{#1}}
\newcommand{\cre}[1]{\hat{c}^\dagger_{#1}}

\begin{document}

\title{Phase Diagram of Mixed-Dimensional Anisotropic $t-J$-Models}
\author{Julius Dicke}
\affiliation{Arnold Sommerfeld center for theoretical physics, LMU Munich, D-80333 Munich, Germany}
\affiliation{Munich Center for Quantum Science and Technology (MCQST), Schellingstr. 4, 80799 M\"unchen, Germany}

\author{Lukas Rammelm\"uller}
\affiliation{Arnold Sommerfeld center for theoretical physics, LMU Munich, D-80333 Munich, Germany}
\affiliation{Munich Center for Quantum Science and Technology (MCQST), Schellingstr. 4, 80799 M\"unchen, Germany}

%\author{{\color{red} more authors}}
% \affiliation{Arnold Sommerfeld center for theoretical physics, LMU Munich, D-80333 Munich, Germany}

\author{Fabian Grusdt}
\affiliation{Arnold Sommerfeld center for theoretical physics, LMU Munich, D-80333 Munich, Germany}
\affiliation{Munich Center for Quantum Science and Technology (MCQST), Schellingstr. 4, 80799 M\"unchen, Germany}

\author{Lode Pollet}
\affiliation{Arnold Sommerfeld center for theoretical physics, LMU Munich, D-80333 Munich, Germany}
\affiliation{Munich Center for Quantum Science and Technology (MCQST), Schellingstr. 4, 80799 M\"unchen, Germany}

\begin{abstract}
We study the phase diagram of two different mixed-dimensional $t-J_z-J_{\perp}$-models on the square lattice, in which the hopping amplitude $t$ is only nonzero along the $x$-direction. In the first model, which is bosonic, the spin exchange amplitude $J_{\perp}$ is negative and isotropic along the $x$ and $y$ directions of the lattice, and $J_z$ is isotropic and positive. The low-energy physics is characterized by spin-charge separation: the holes hop as free fermions in an easy-plane ferromagnetic background. In the second model, $J_{\perp}$ is restricted to the $x$-axis while $J_z$ remains isotropic and positive. The model is agnostic to particle statistics, and shows stripe patterns with anti-ferromagnetic N{\'e}el order at low temperature and high hole densities, in resemblance of the mixed-dimensional $t-J_z$ and $t-J$ models. At lower hole density, a very strong first order transition and hysteresis loop is seen extending to a remarkably high 14(1)\% hole doping.
\end{abstract}
\maketitle

%%%%%%%%%%%%%%%%%%%%%%%%%%%%%%%%%%%%%%%%%%%%%%%%%%%%%%%%%%
%%%%%%%%%%%%%%%%%%%%%%%%%%%%%%%%%%%%%%%%%%%%%%%%%%%%%%%%%%
%%%%%%%%%%%%%%%%%%%%%%%%%%%%%%%%%%%%%%%%%%%%%%%%%%%%%%%%%%
\section{Introduction}

Competing orders, in which radically different phases appear with very similar ordering temperatures, are a hallmark of highly correlated systems and high-temperature superconductors in particular~\cite{Dagotto1994_RMP, Lee2006_RMP, Kivelson2003_RMP, Fradkin2015_RMP}. The interplay between hole motion and local antiferromagnetic correlations plays a crucial role, although antiferromagnetism and superconductivity are rarely seen together and tend to suppress each other on long length scales.
The most recent studies find stripe order in the Hubbard and $t-J$ models~\cite{Corboz2014, SimonsCollaboration_2015, Zheng2016, Qin2020}, which are the minimal models capturing the localization of the $d-$electrons. But it has also been known that various ground states of different nature may be close in energy~\cite{White1997}, with extreme sensitivity to parameters such as the next-nearest hopping amplitude $t'$~\cite{Jiang2020}.  Whereas it is difficult to estimate the role of an individual parameter in a frustrated parameter landscape, nematicity is generally accepted to favor stripe formation~\cite{Kivelson2003_RMP}.

In previous work, two of us studied the phase diagram of the mixed-dimensional $t-J_z$-model~\cite{Grusdt2021}, in which the hopping is one-dimensional (along $x$) but the Ising spins interact via $J_z$ isotropically and anti-ferromagnetically. Its phase diagram showed a chargon phase ({\it i.e.}, a phase without order) at high temperature, stripes with antiferromagnetic domains modulated by the hole doping at low temperature, and a meson phase consisting of paired holes at low hole density and at low but finite temperature. The phase diagram had a natural explanation in terms of a $\mathbb{Z}_2$ lattice gauge theory. The purpose of the present paper is to examine the effect of quantum spin exchange interactions absent in Ref.~\cite{Grusdt2021}. This will be done for two different cases, (i) spatially isotropic but ferromagnetic spin-exchange interactions (a bosonic model) realized for $\alpha = 1$ in Eq.~\ref{eq:hamiltonian}, (ii) spatially anisotropic spin-exchange interactions purely oriented along $x$ realized for $\alpha = 0$ in Eq.~\ref{eq:hamiltonian}.

The second purpose of our work is to study how a one-dimensional hopping amplitude affects the results found for isotropic hopping amplitudes and ferromagnetic spin-exchange interactions, as was studied in the bosonic $t-J_z-J_{\perp}$-model of Ref.~\cite{Boninsegni2006,Boninsegni2008}. The authors found for any $J_{\perp} / J_z < 1$ phase separation into a superfluid hole-rich and an antiferromagnetic hole-free region at low enough hole density. Its occurrence was claimed on the basis of the string picture of Brinkman and Rice~\cite{BrinkmanRice_1970}, which excludes the case $J_{\perp} = J_z$~\cite{Boninsegni1992}.
The physics of models interpolating between the one of Refs.~\cite{Boninsegni2006,Boninsegni2008} on the one hand and Ref.~\cite{Grusdt2021} on the other hand, will be reported here.

Our main result is the description of all phases for the models with $\alpha = 1$ and $\alpha = 0$ in Eq.~\ref{eq:hamiltonian} at constant $\vert J_{\perp} \vert  = J_z = 0.4$ (the in-plane component is ferromagnetic), and hopping amplitude $t=1$ for temperatures down to $\beta t = 20$ obtained by first principles quantum Monte Carlo simulations. For $\alpha = 1$, the spins show an easy-plane ferromagnetic order for non-zero hole doping. The holes move in this spin background as if they were free particles, showing Friedel oscillations. For $\alpha = 0$, we find at large hole doping density $n_h > 14(1)$\% a N{\'e}el-type, hidden antiferromagnetic order. The holes form a periodic structure along the $x-$direction, leading to vertical stripes. The spins are oriented in opposite directions across a hole. The N{\'e}el temperature and the hole ordering temperature coincide, making the transition first order. At lower hole densities, we see clear signs of a very strong first order transition, accompanied by a strong hysteresis loop, between the  hole-free perfect N{\'e}el state and a phase with the spin and charge stripes described above.

The paper is structured as follows. In Sec.~\ref{sec:model} we introduce the model and define the main observables of interest. In Sec.~\ref{sec:isotropic} we report our findings for the model with $\alpha=1$, which lead to a picture of an effective spin-charge separation. In Sec.~\ref{sec:anisotropic} we show how vertical stripes for large hole densities and hysteresis for low hole densities are found in the model with $\alpha = 0$. We conclude in Sec.~\ref{sec:conclusion}.

%%%%%%%%%%%%%%%%%%%%%%%%%%%%%%%%%%%%%%%%%%%%%%%%%%%%%%%%%%
%%%%%%%%%%%%%%%%%%%%%%%%%%%%%%%%%%%%%%%%%%%%%%%%%%%%%%%%%%
%%%%%%%%%%%%%%%%%%%%%%%%%%%%%%%%%%%%%%%%%%%%%%%%%%%%%%%%%%
\section{Model}
\label{sec:model}
The system is defined on the square lattice of size $L_x \times L_y$ with lattice constant $a=1$  by the Hamiltonian

\begin{align} 
  \label{eq:hamiltonian}
  \hat H = &-t\sum_\sigma\sum_{\ev{\bf i,j}_{\hat{x}}} \left( \mP\cre{{\bf i},\sigma} \ani{{\bf j},\sigma}\mP + {\rm h.c.} \right) \nonumber \\
  & +  J_z \sum_{\ev{\bf i,j}} \left(\sz{{\bf i}}\sz{{\bf j}} - \frac{\hat{n}_{{\bf i}}\hat{n}_{{\bf j}}}{4}\right) - \frac{J_\perp}{2} \sum_{\ev{\bf i,j}_{\hat{x}}} \left(  \sp{{\bf i}}\sm{{\bf j}} +{\rm h.c.} \right) \nonumber \\
  & -\alpha\frac{J_\perp}{2} \sum_{\ev{\bf i,j}_{\hat{y}}} \left( \sp{{\bf i}}\sm{{\bf j}} +{\rm h.c.} \right),
\end{align}
where $\cre{{\bf i},\sigma}$ creates a boson of type $\sigma$ at site ${\bf i}$ and correspondingly $\ani{{\bf j},\sigma}$ destroys one. All parameters are taken positive. The notation $\ev{\bf i,j} (\ev{\bf i,j}_{\hat{\mu}})$ denotes a pair of neigboring sites $\mathbf{i}$ and $\mathbf{j}$ (along direction $\hat{\mu} = \hat{x}, \hat{y}$). \\

The operator $\hat{n}_{\bf i} = \sum_{\sigma}  \cre{{\bf i},\sigma}\ani{{\bf i},\sigma} $ counts the total number of particles on lattice site $\bf{i}$. The operators $\sz{{\bf i}} = \frac{1}{2} \left( \hat{n}_{i, \uparrow} - \hat{n}_{i, \downarrow} \right) \, , \, \sp{{\bf i}} = \cre{{\bf i},\uparrow} \ani{{\bf i},\downarrow}$, and $ \sm{{\bf i}} = \cre{{\bf i},\downarrow} \ani{{\bf i},\uparrow}$ are the usual spin-$1/2$ operators.
 The projectors $\mP$ ensure that at most one particle resides on a given lattice site. We consider two different cases: (i) fully isotropic spin-exchange interaction ($\alpha=1$),  and (ii) spin-exchange terms restricted to the $x$-direction only ($\alpha=0$). Unless otherwise noted, all results will be presented for the isotropic case $J_z = J_\perp = 0.4$ with hopping set to $t=1$, such that $t/J = 2.5$ (corresponding to $U = 4 t^2/J =10$ in terms of a Hubbard model). This is also a typical value for the cuprates.

\begin{figure}
  \includegraphics[width=\columnwidth]{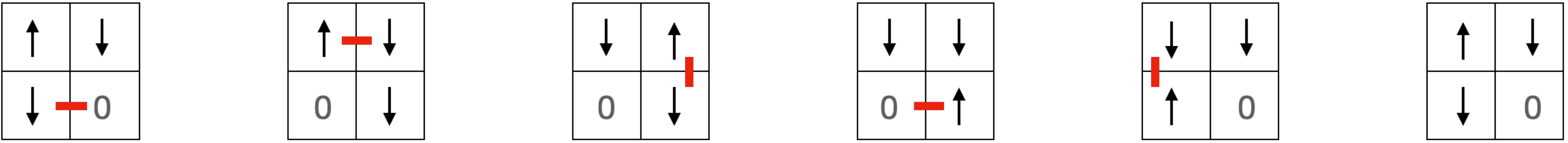}
  \caption{\label{fig:sign_loop} Illustration of an exchange cycle that leads to a sign problem for in-plane antiferromagnetic ({\it i.e.},  $J_{\perp} < 0$ in Eq.~\ref{eq:hamiltonian}) spin coupling when $\alpha > 0$. The red line indicates which spins or holes are exchanged in each step. The loop consists of moving the hole once and then moving the up-spin clockwise around the plaquette. The last configuration is the same as the initial configuration because of the indistinguishability of the down spins. The shown exchange loop involves two hopping amplitudes, two spin exchanges in the y-direction and one in the x-direction. It is the latter that is relevant for the sign structure, even on a bipartite lattice, since the sign of the configuration is given by $(-1)^{n_{\rm hop} + n_{\rm s.e.}}$, with $n_{\rm hop}$ and $n_{\rm s.e.}$ the total number of particle hopping and spin exchanges, respectively, present in the path integral configuration.
    }
\end{figure}

The model in \equref{hamiltonian} is written such that positive $J_z$ reflects antiferromagnetic (AFM) couplings in the $z$-direction but positive $J_\perp$ reflects in-plane ferromagnetic (FM) coupling, which is necessary to prevent the sign problem for $\alpha >0$, as is explained in Fig.~\ref{fig:sign_loop}. For $\alpha = 0$, the model is sign-free and furthermore particle statistics agnostic in case of open boundary conditions since the charge degrees of freedom never leave a row. The model possesses spin $SU(2)$ symmetry only at zero hole density and if simultaneously $\alpha=1$.

Regardless of the nature of the spin-flipping term, the restricted 1D hopping in \equref{hamiltonian} conserves the number of holes per row. We always choose the same number of holes per row in canonical simulations, or the same chemical potential in grand-canonical simulations.
Moreover, for $\alpha=0$, the $\up$- and $\down$-densities are conserved separately in each row. We usually work in the zero magnetization sector for an even number of holes, and in the sector with total $S^z = 1/2$ for an odd number of holes.
The model is addressed with  path integral quantum Monte Carlo (QMC) simulations using worm updates~\cite{Prokofev_1998}  in a straightforward adaptation of Ref.~\cite{Pollet2007}. 
%We employ periodic boundary conditions -- this is a difference to the problems studied in Ref.~\cite{Grusdt2021} and all DMRG studies, which employ open boundary conditions.
The main quantities of interest are the kinetic, potential, and total energy, the spin-spin correlation functions $C_z(\mathbf{r}_1, \mathbf{r}_2) = \left< S^z_{\mathbf{r}_1} S^z_{\mathbf{r}_2} \right>$ and $C_{\perp}(\mathbf{r}_1, \mathbf{r}_2) = \left< S^+_{\mathbf{r}_1} S^-_{\mathbf{r}_2 }+ {\rm h.c.} \,\, \right>$, and the hole-hole correlation function $\rho_{hh} (\mathbf{r}_1, \mathbf{r}_2) = \left< (1 - n_{\mathbf{r}_1})(1 - n_{\mathbf{r}_2}) \right>$. For translationally invariant systems we also use the notation $C_c(\mathbf{r}) = C_c(\mathbf{r} = \mathbf{r}_1 - \mathbf{r}_2, 0)$, with $c = z, \perp$.

Experimentally, the model in Eq.~\ref{eq:hamiltonian} can be realized starting from pseudo-spin-$1/2$ bosons on a square lattice and subject to a strong tilt $\Delta$ along the $\hat{y}-$direction in order to inhibit particle tunnelling along $\hat{y}$~\cite{Grusdt2018}. For large on-site Hubbard interactions, the bosonic superexchange along $\hat{y}$ realizes ferromagnetic terms on $\ev{\bf i,j}_{\hat{x}}$, with $J_{\perp} > 0$ in Eq.~\ref{eq:hamiltonian}~\cite{Duan2003,Trotzky2008}. Similarly, superexchange along the gradient realizes $J_{\perp}$-terms with tunable sign on $\ev{\bf i,j}_{\hat{y}}$~\cite{Trotzky2008, Hirthe2022}. By adding spatially isotropic Rydberg-Rydberg interactions the Ising terms in Eq.~\ref{eq:hamiltonian} can be realized~\cite{Zeiher2016,GuardadoSanchez2021}. For $\alpha = 0$, the gradient $\Delta$ can be removed if a sufficiently strong anisotropic lattice guarantees that tunnelling along $\hat{y}$ can be neglected, $t_y = 0$ along $\ev{\bf i,j}_{\hat{y}}$ as required.

\section{Model with isotropic spin-exchange terms $\alpha = 1$ }
\label{sec:isotropic}

\begin{figure}
  \centering
  \includegraphics[width=0.9\columnwidth]{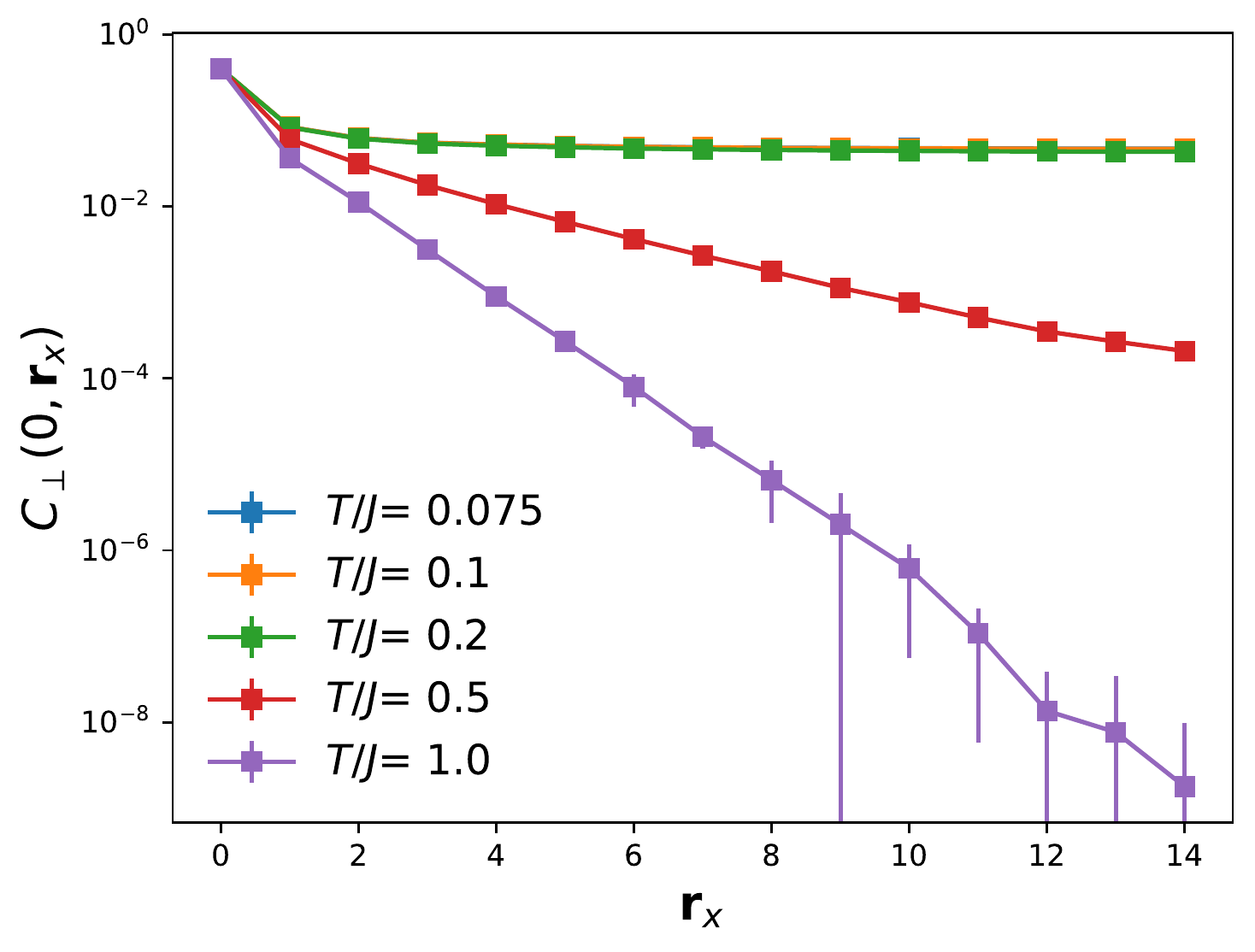}
  \caption{\label{fig:xy_2dflips} 
  In-plane spin ordering for the model  with $\alpha = 1$ on a lattice of size $L_x\times L_y = 30\times 30$ with periodic boundary conditions and hole density $n_h = 6/30$ per row. The plot shows the spatial dependence of the the easy-plane correlation function $C_{\perp}( \mathbf{r} )$ for several temperatures. The curves for the three lowest temperatures lie on top of each other, where near-ordering is seen because of the finite system size. The correlator decays as a powerlaw at any finite temperature in the thermodynamic limit.
  }
\end{figure}

\begin{figure*}[t]
    \includegraphics[width=\textwidth]{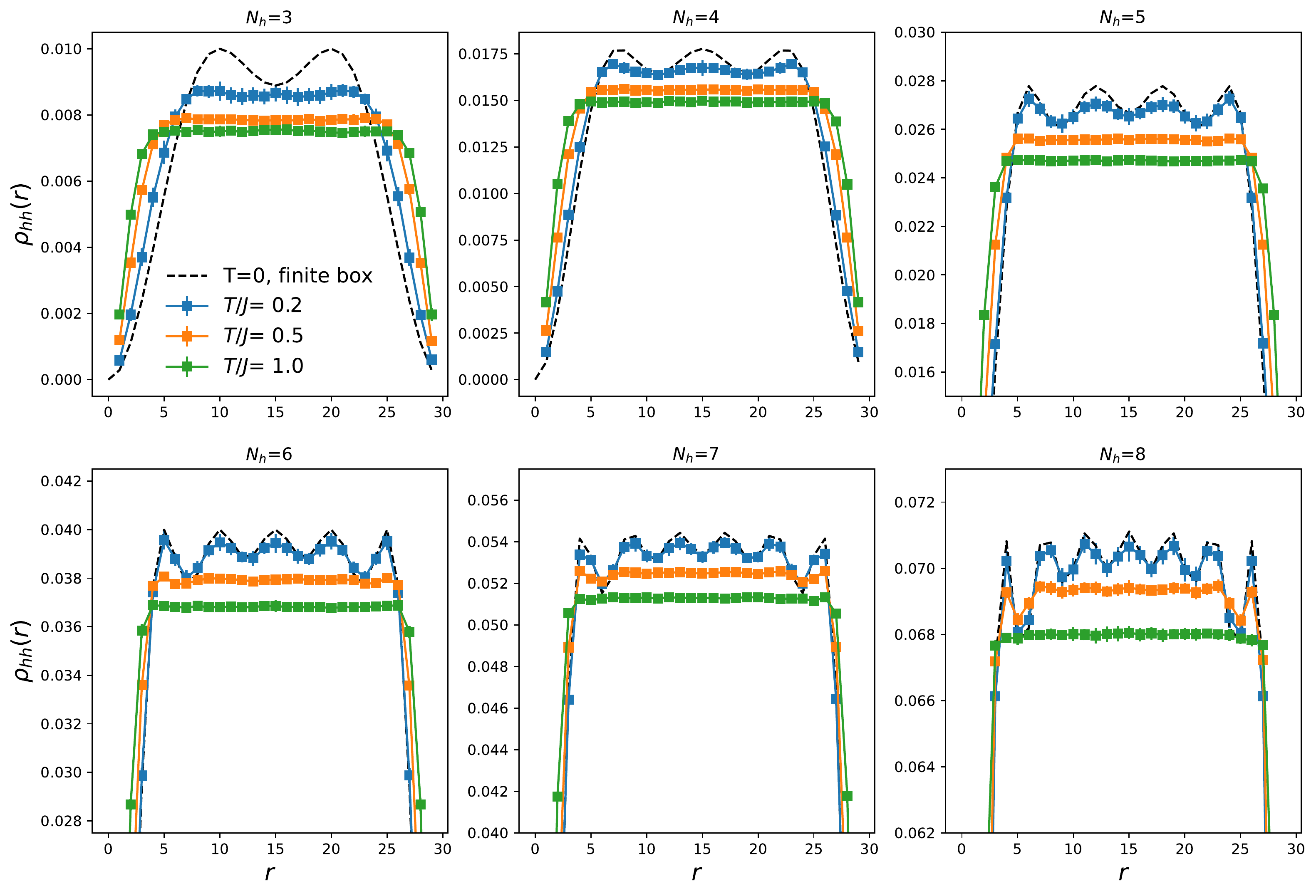}
    \caption{Hole-hole correlation function along the $x$-direction for hole densities $n_h = 3/30, 4/30, 5/30, 6/30, 7/30, 8/30$ (top left to bottom right). The dashed black line corresponds to the expression of non-interacting 1D spinless fermions at zero temperature. Periodic boundary conditions are used }
    \label{fig:hole_hole_corr}
\end{figure*}

For the spin-isotropic mixed-dimensional $t-J$ model with $\alpha = 1$ in \equref{hamiltonian}, no simple $\mathbb{Z}_2$ lattice gauge structure can be introduced, unlike Ref.~\cite{Grusdt2021}.
We see in Fig.~\ref{fig:xy_2dflips} that the spins show ferromagnetic easy-plane ordering at low temperature for any nonzero hole doping: the $C_{\perp}( \mathbf{r} )$ correlator approaches a temperature-independent constant at large distances on our finite system size. 
Spin correlations along the $z$-axis decay exponentially, unless at zero hole density when the spin $SU(2)$ symmetry is preserved.  For finite hole density, the spin symmetry is $U(1)$. Since it is impossible to break a continuous symmetry in two dimensions at finite temperature thanks to the Mermin-Hohenberg-Wagner theorem~\cite{Mermin1966,Hohenberg1967}, true long-range order cannot exist. However, when lowering the temperature from the paramagnetic phase, a Kosterlitz-Thouless transition to a quasi-long range order is nevertheless  possible~\cite{Kosterlitz1973}. Such a quasi-long range ordered phase has therefore a zero in-plane magnetization, but  a non-zero spin stiffness. This can be detected in the path integral Monte Carlo simulations by computing the winding number fluctuations, which are proportonal to the spin stiffness~\cite{Pollock_1987}.
%and the transition we observe is of the Kosterlitz-Thouless type with a critical temperature that is furthermore dependent on the hole density. Note that in the thermodynamic limit at any finite temperature only quasi-long range order can exist and that at zero temperature true long-range order is established.
%The particle winding number squared, proportional to the spin stiffness~\cite{Pollock_1987}, is non-zero along $x$ below the critical temperature. 
In the charge sector, we observe no correlations between the rows, ruling out any type of stripe order. Within each row however, Friedel oscillations in the hole-hole correlation function  can be observed at low enough temperature, as is shown in Fig.~\ref{fig:hole_hole_corr}. The density-density correlation function can be computed analytically from $\rho^0(r,0) \rho^0(0, r)$, where the one-body density matrix for spinless free fermions is given by $\rho^0(r,r') = \frac{1}{L_x}\sum_k n_k \e^{\i k (r-r')}$. 
Here, $n_k$ is the Fermi-Dirac distribution, $n_k = \frac{1}{\exp(\beta \epsilon_k) + 1}$ with $\epsilon_k = -2t \cos k - \mu$, $\mu$ the chemical potential, and $k = \frac{2 \pi n_x}{L_x}, n_x = 0, 1, \ldots L_x-1$. In \figref{hole_hole_corr} we used the $T=0$ expression where the momentum distribution simplifies to a step-function with the jump at $k_F$.  Note that fewer holes require lower temperatures before the ground state is reached.
Similar to the charge- and spin-isotropic case considered in Refs.~\cite{Boninsegni2006,Boninsegni2008} we observe no indication of phase separation for $\alpha=1$. 

Our observations can be explained in terms of a generalized spin-charge separation picture. Although spin-charge separation is well established for one-dimensional systems, the present system is one-dimensional only in the charge sector.  
Nevertheless, when the spins are ferromagnetically in-plane ordered, there is no penalty for moving the holes around, unlike in the string picture for antiferromagnetic order, and the spin and charge sectors can hence effectively decouple. We recall that an effective decoupling was also observed in the different model of Ref.~\cite{Gukelberger2017}, where the hopping amplitudes were  one-dimensional but in orthogonal directions for up and down spins.

%%%%%%%%%%%%%%%%%%%%%%%%%%%%%%%%%%%%%%%%%%%%%%%%%%%%%%%%%%
%%%%%%%%%%%%%%%%%%%%%%%%%%%%%%%%%%%%%%%%%%%%%%%%%%%%%%%%%%
%%%%%%%%%%%%%%%%%%%%%%%%%%%%%%%%%%%%%%%%%%%%%%%%%%%%%%%%%%
\section{Model with isotropic spin-exchange terms $\alpha = 0$}
\label{sec:anisotropic}

We proceed with the model in which spin-exchange loops are turned off: The only closed loops one can make with the hopping and spin-exchange terms in the Hamiltonian for $\alpha = 0$ are backtracking ones (the loops shown in Fig.~\ref{fig:sign_loop} are impossible), rendering the system agnostic to the particle statistics. The two-dimensional nature of the $J_z$ term does not alter this but is expected to enhance easy-axis anti-ferromagnetism, similar to the physics of the mixed-dimensional $t-J_z$-model of Ref.~\cite{Grusdt2021}. We thus focus on stripe formation with anti-ferromagnetic easy-axis order. At low hole doping however, the antiferromagnetic order gives rise to a string picture~\cite{BrinkmanRice_1970} when moving holes around. As this leads to a strong increase in the potential energy that cannot be fully mended due to the lack of quantum fluctuations  along the $y$-direction, we investigate a possible phase separation.

\subsection{The regime of low hole density}\label{sec:phasesep}

\begin{figure}
  \includegraphics[width=\columnwidth]{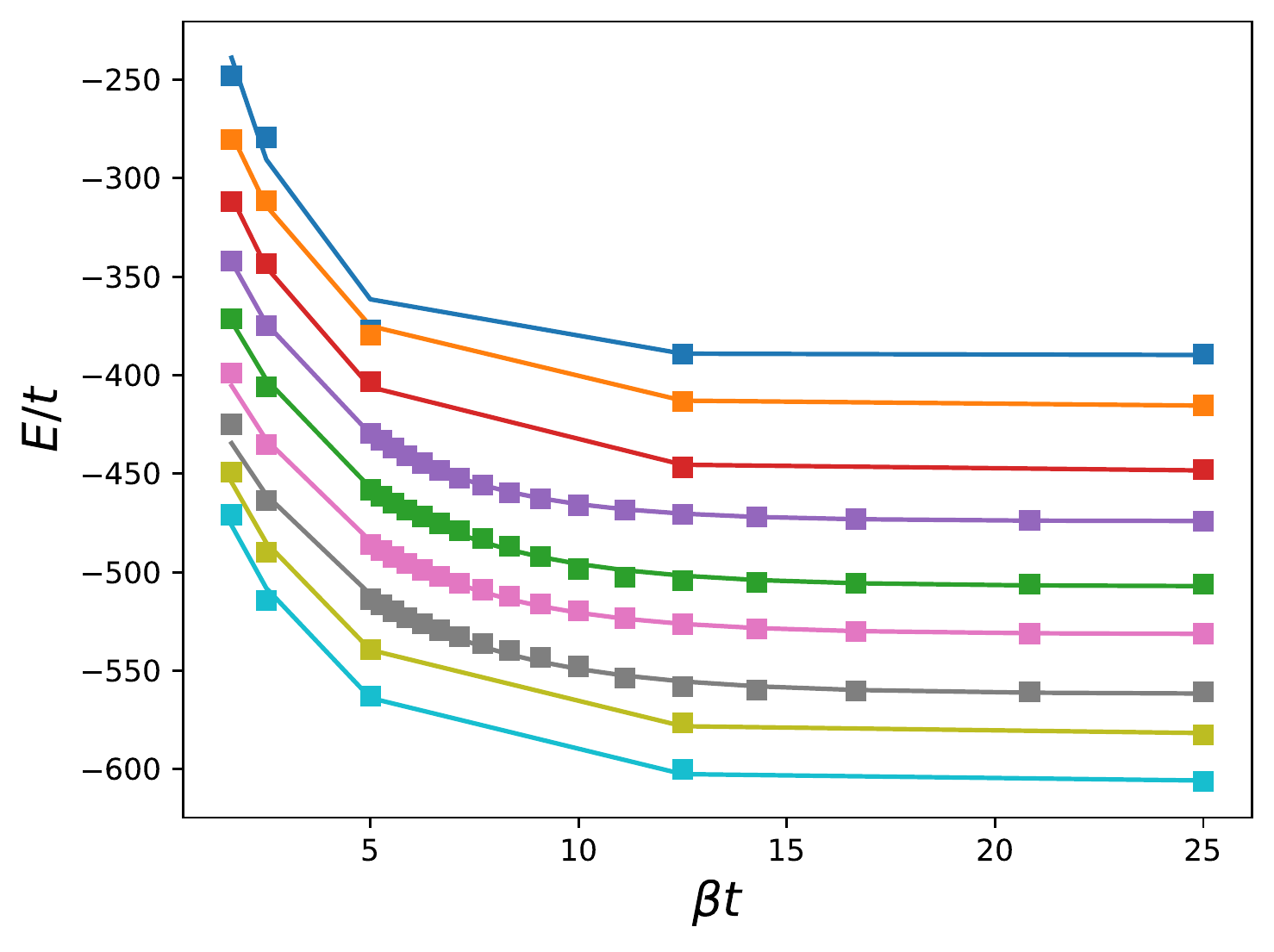}
  \caption{\label{fig:phase_separation1} Extrapolation of the total energy to the ground state for various hole densities for a system of $L_x\times L_y = 30\times 30$ lattices with periodic boundary conditions. The curves correspond to different numbers of holes per row, ranging from 0 to 8, top to bottom.  }
\end{figure}

\begin{figure}
  \includegraphics[width=\columnwidth]{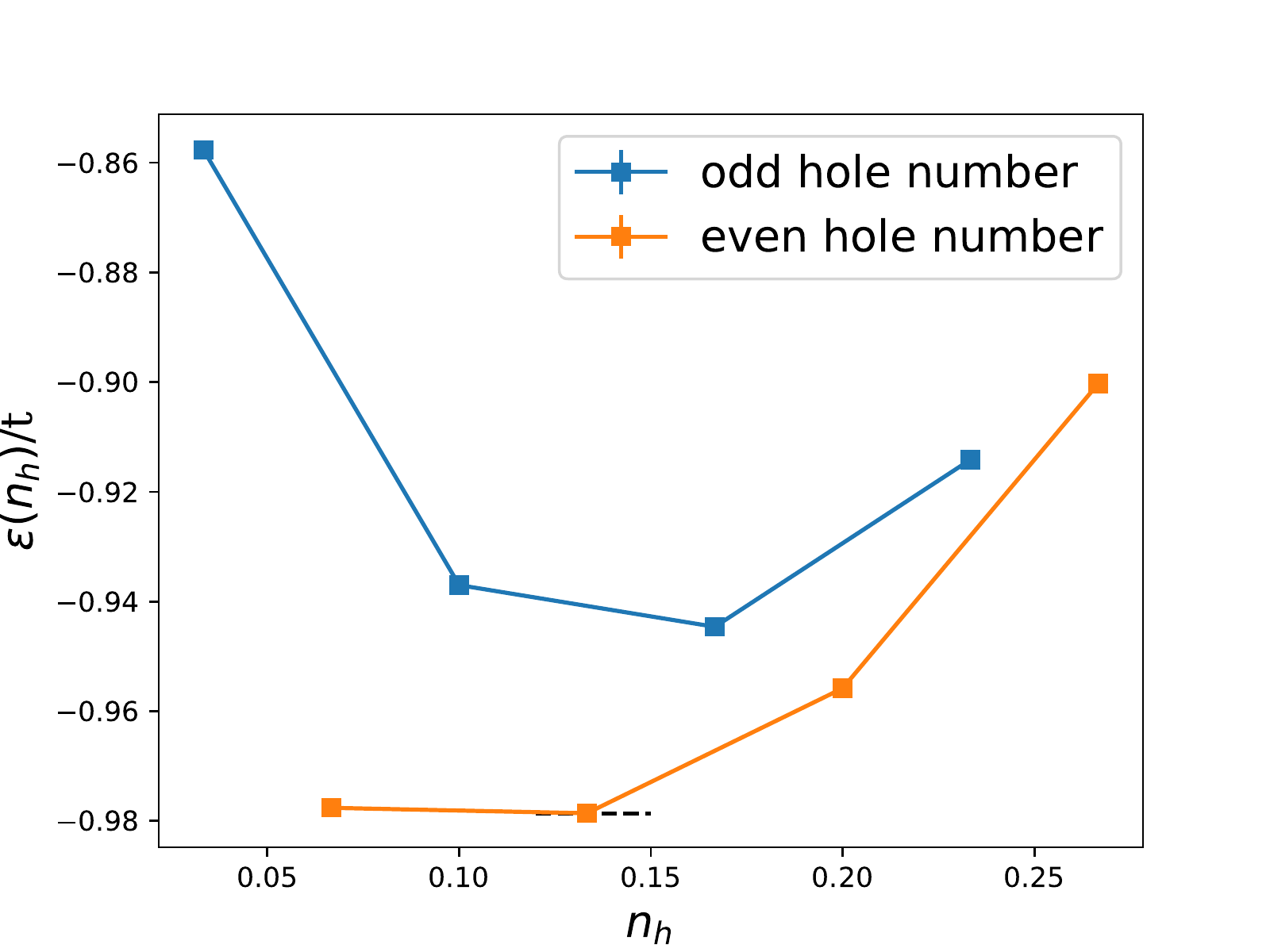}
  \caption{\label{fig:phase_separation}  Energy per hole (Eq.~\ref{eq:e_hole}) at zero temperature as a function of hole density extracted from Fig.~\ref{fig:phase_separation1}. Odd and even hole numbers are plotted separately but only the even hole numbers are relevant. The lowest energy is indicated by the black dashed line at a hole density of $4/30 \approx 0.1333$.
   }
\end{figure}

\begin{figure}
  \includegraphics[width=\columnwidth]{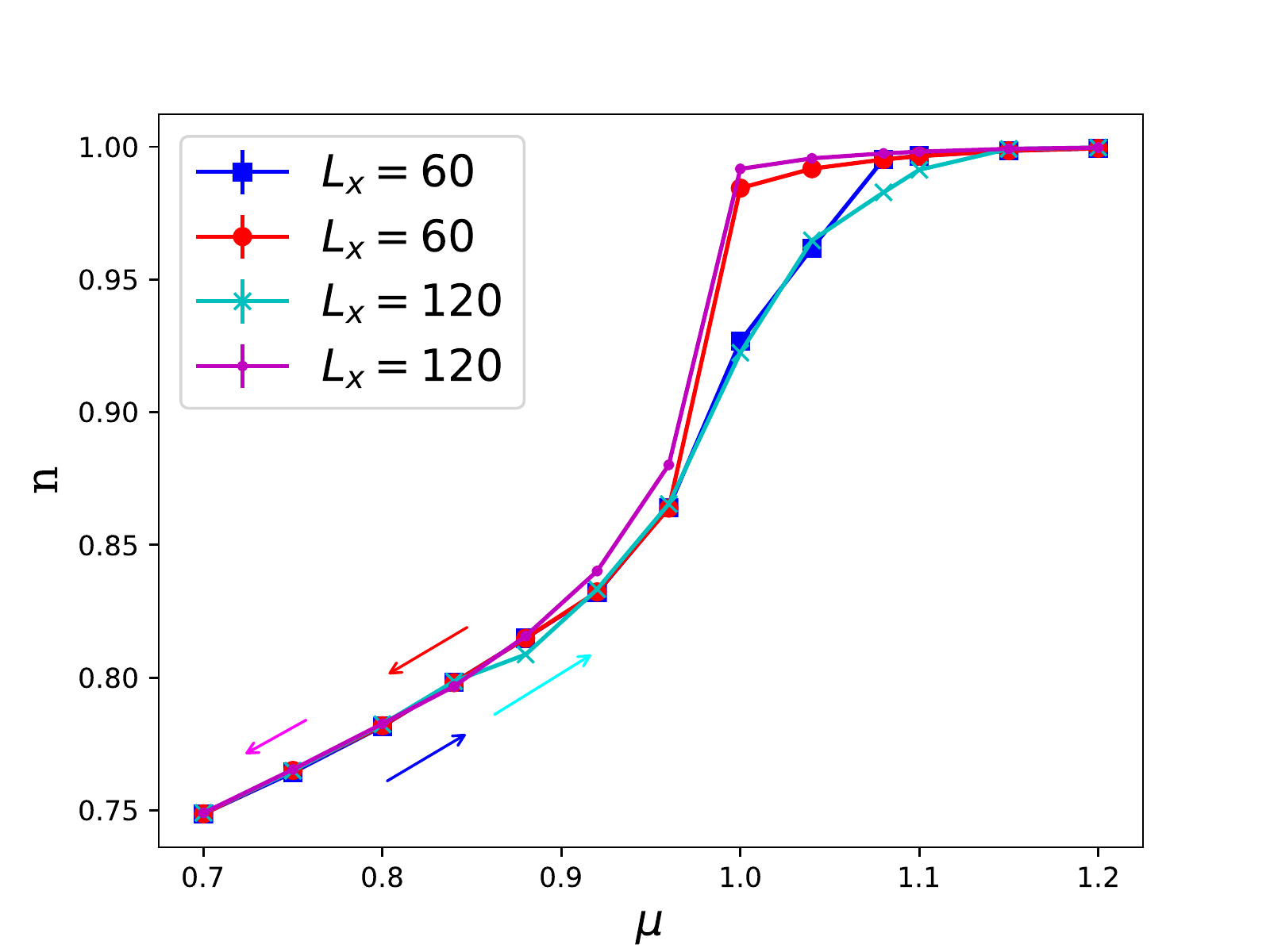}
  \caption{\label{fig:hysteresis} Particle number as a function of chemical potential $\mu$ (taken constant for each row) for a system with parameters $t=1, J=0.4, \alpha=0, \beta=20, L_y=L_x/2$, open boundary conditions and two different system sizes, $L_x=60$ and $L_x=120$. The arrows indicate the hysteresis loop; each value of $\mu$ was initiated with a well-annealed sample from the previous $\mu$-value along the loop.  
  }
\end{figure}

Phase separation for low hole doping in the (fermionic) $t-J$-models has a long history and remains a controversial issue in two dimensions for physically relevant values of $J$~\cite{Emery1990, Puttika1992, Hellberg1996, Calandra1998, Bachelet1998, Hellberg2000, Rommer2000, Shih2001, Scalapino2012}. But for quasi-1D systems consisting of ladders with just a few legs, DMRG studies could clearly establish the phase separation boundary, even at low hole hoping~\cite{Rommer2000, Scalapino2012}: the boundary is close to $J/t = 2$ for 2 legs, and shifts to lower values of $J/t \sim 1.4$ when increasing the number of legs to 6.

In order to examine whether our system phase separates into a hole-rich phase and a hole-free phase,  we follow the argument of Ref.~\cite{Emery1990, Rommer2000}, in the canonical ensemble.
Suppose phase separation is found between a region with zero holes and a hole-rich phase with hole density $x = x_{\rm ps}$ (note that we use the notation $n_h$ for hole densities in simulations, which are restricted to be rational numbers). Let the respective energies be $E(0)$ and $E(x_{\rm ps})$.
Then for any $0<x<x_{\rm ps}$ the energy of the homogeneous phase $E(x)$ will be higher than the linearly interpolated one, $E_{\rm lin}(x) = E(0) + x \frac{E(x_{\rm ps}) - E(0)}{x_{\rm ps}}$. This is just the standard Maxwell argument. 
In the phase separated regime, we will thus measure energies that are linear in the hole concentration in the simulations, provided that the simulations are ergodic and the system sizes sufficiently large (such that both phases fit onto the total system size and the surface energy negligible). If this is not the case, then one might observe a uniform metastable state, with higher energy.
%The energy $E$ is written as  $ E  =  (N_s - N)\epsilon_H + N\epsilon_h(n_h) $,  where $N_s$ is the total number of sites, $N$ the number of sites in the hole-rich phase, $\epsilon_H$  the energy per site in the hole-free region, and $\epsilon_h(n_h)$ the energy per site in the hole-rich phase.  If $E$ has a minimum as a function of the hole density  $n_h = N_h/N$ for $n_h > 0$, then it is favorable for the system to phase separate; {\it i.e.}, phase separation occurs when the function,
In order to find the density $x_{\rm ps}$ of the hole-rich phase, one can directly use the previous formula by minimizing $E_{\rm lin}(x)$ with respect to $x_{\rm ps}$. It follows that the linear curve is also tangential to $E(x)$ at $x=x_{\rm ps}$.
Our energies per site were found to be quasi-linear in the doping (not shown), but with an upward curvature clearly seen for 6 holes (with linear size $L_x = 30$, thus $n_h = 0.2$). Since matching the tangential slope is difficult, we switched to the original argument by Emery et al.~\cite{Emery1990} to determine $x_{\rm ps}$, and which was for instance also used in Ref.~\cite{Bachelet1998, Shih2001, Boninsegni2008}. It is likewise based on the existence of a negative inverse compressibility for a finite system~\cite{Bachelet1998}. One determines the minimum of the energy  per hole, which is defined as
\begin{equation}
  \label{eq:e_hole}
  %\epsilon(n_h) = \frac{\epsilon_h(n_h) - \epsilon_H}{n_h},
  \epsilon(x) = \frac{E(x) - E(0)}{N_x},
\end{equation}
where $N_x$ is the total number of holes. The hole density for which the minimum is found, is $x_{\rm ps}$. We expect this formula to work well when $x_{\rm ps}$ is not too small (as it enters the denominator in Eq.~\ref{eq:e_hole}). For hole densities $x < x_{\rm ps}$ the system must phase separate: it is energetically favorable to create a region with hole density $x_{\rm ps}$ and one without holes in such a way that the total hole concentration remains $x$.
The energy of the separated state is lower than the one of the uniform state.

In \figref{phase_separation1} we show the extrapolation of the total energy on a $30\times 30$ lattice to zero temperatures. From \equref{e_hole} we obtain then the energy per hole, shown in \figref{phase_separation}. We observe a strong odd/even effect, owing to the presence of geometric frustration in the case of odd hole numbers with periodic boundary conditions. The curve for even hole numbers (the one relevant for us) shows a flattening at low hole density with a shallow minimum for a hole density $n_h = 4/30$. %Given the flatness of the curve, it is a good idea to have a second calculation to support this claim.

In order to have a different view on the problem of phase separation, we switch to grand-canonical simulations and found strong indications of a hysteresis loop at low hole doping, as is shown in Fig.~\ref{fig:hysteresis}. 
A hysteresis loop is a characteristic of a strong first order transition, in this case between an anti-ferromagnetic defect-free state, and a striped phase that is discussed further in the next section. 
The hysteresis loop persists to a remarkably high hole density in line with the value of the previous parapraph, $n_h \approx 0.14$ (for $L_x = 60$), and is only seen at low temperature: The figure was taken for $\beta t = 20$ but we saw no indications of hysteresis for $\beta t = 10$ (as we will see in the next section, the critical temperature for stripe formation is also lower than $\beta t = 10$).
For slightly higher hole densities  in the range $ 0.15 \le n_h \le 0.25$, simulations converge slowly, with plenty of metastable states. For $L_x = 60$ however, the Monte Carlo algorithm was eventually able to tunnel out of the metastable minima. For $L_x = 120$ this was not always the case, as can be seen in Fig.~\ref{fig:hysteresis} (see eg the data point at $n_h = 0.88$), and also the hysteresis loop did not fully close. The latter can be made plausible  for open boundary conditions as follows: For $\mu/t \ge 1$ the system on the upper branch is for $L_x = 120$ closer to a hole-free antiferromagnet than for $L_x = 60$, since the holes are found almost exclusively in the lowest and highest rows. Hence, more rows have defect-free antiferromagnetic order for $L_x=120, L_y=60$ than for $L_x=60, L_y=30$.
After the sizeable density jump seen between $\mu/t = 0.96$ and $\mu/t=1$ of the order of $14$\% one would then expect that the densities at $\mu/t = 0.96$, and to a lesser degree at $\mu/t=0.92$, might still differ somewhat between $L_x=60$ and $L_x=120$.

Our observation of phase separation for $J = 0.4$ is hence in line with the bosonic $t-J$ model of Ref.~\cite{Boninsegni2008} but casts little light on the fermionic system.

\subsection{The regime of  large hole density}

\begin{figure}
 \includegraphics[width=\columnwidth]{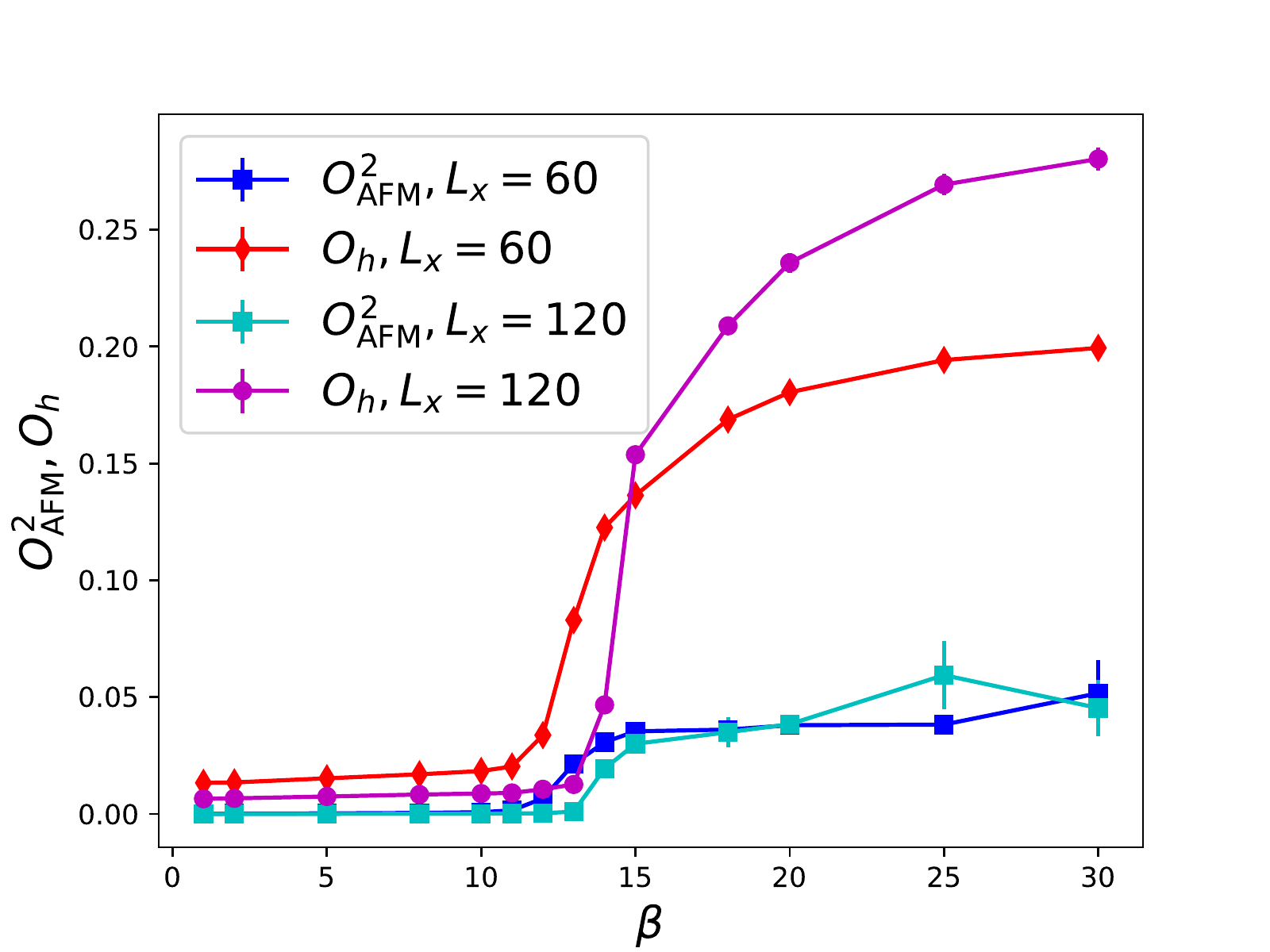}
  \caption{Antiferromagnetic  and hole ordering order parameters for the model with $\alpha=0$ and parameters $t=1, J_{\perp}=J_z=0.4, L_y = L_x/2$, periodic boundary conditions and two different $L_x=60$ and $120$. Simulations are performed in the canonical ensemble with $20$\% hole doping per row ($n_h = 0.2$).
  }
  \label{fig:stripes20}
\end{figure}
\begin{figure}
 \includegraphics[width=\columnwidth]{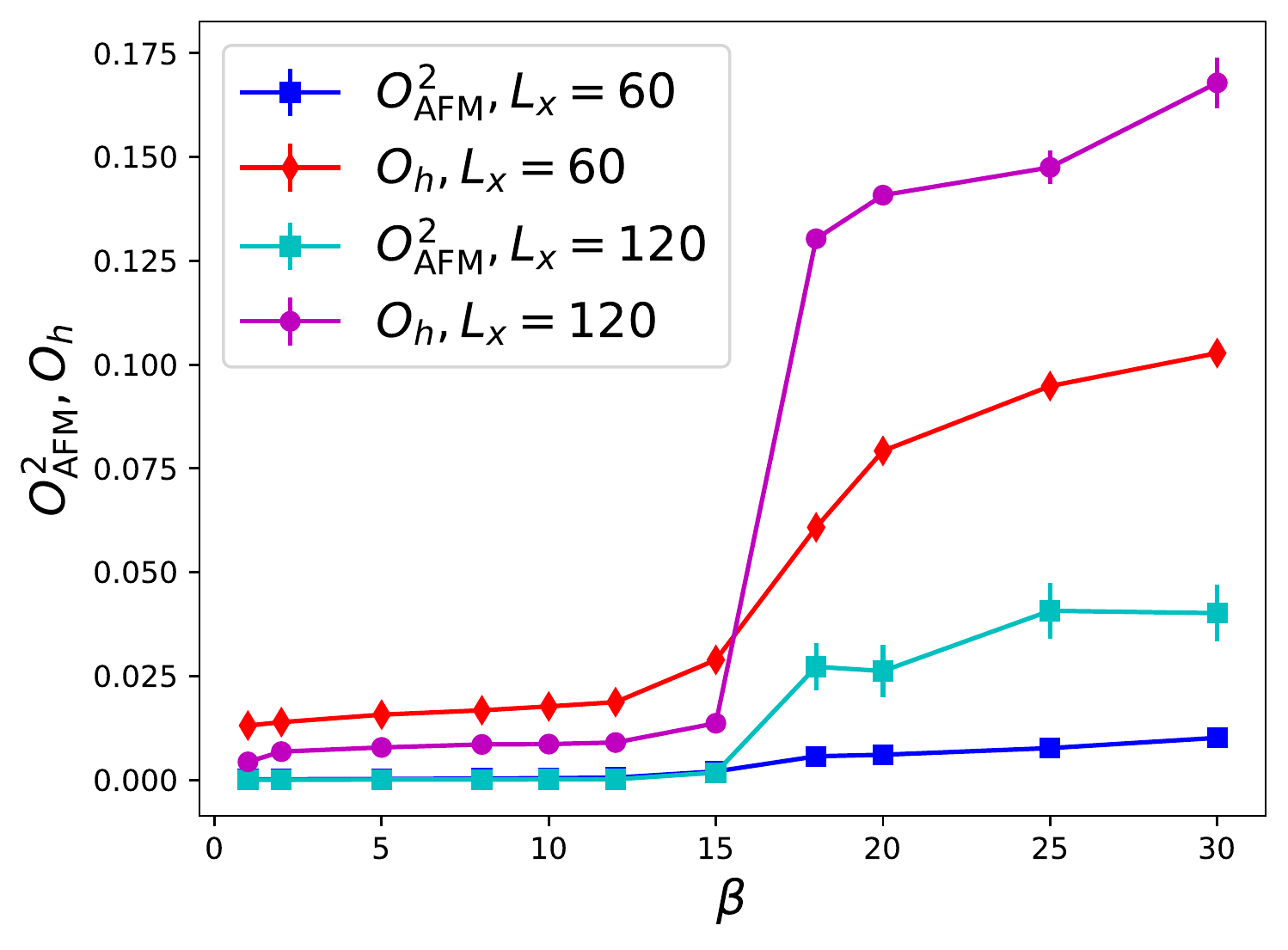}
  \caption{Antiferromagnetic  and hole ordering order parameters for the model with $\alpha=0$ and parameters $t=1, J_{\perp}=J_z=0.4,  L_y = L_x/2$, periodic boundary conditions and two different $L_x=60$ and $120$. Simulations are performed in the canonical ensemble with $25$\% hole doping per row  ($n_h = 0.25$).
  }
  \label{fig:stripes25}
\end{figure}
\begin{figure}
 \includegraphics[width=\columnwidth]{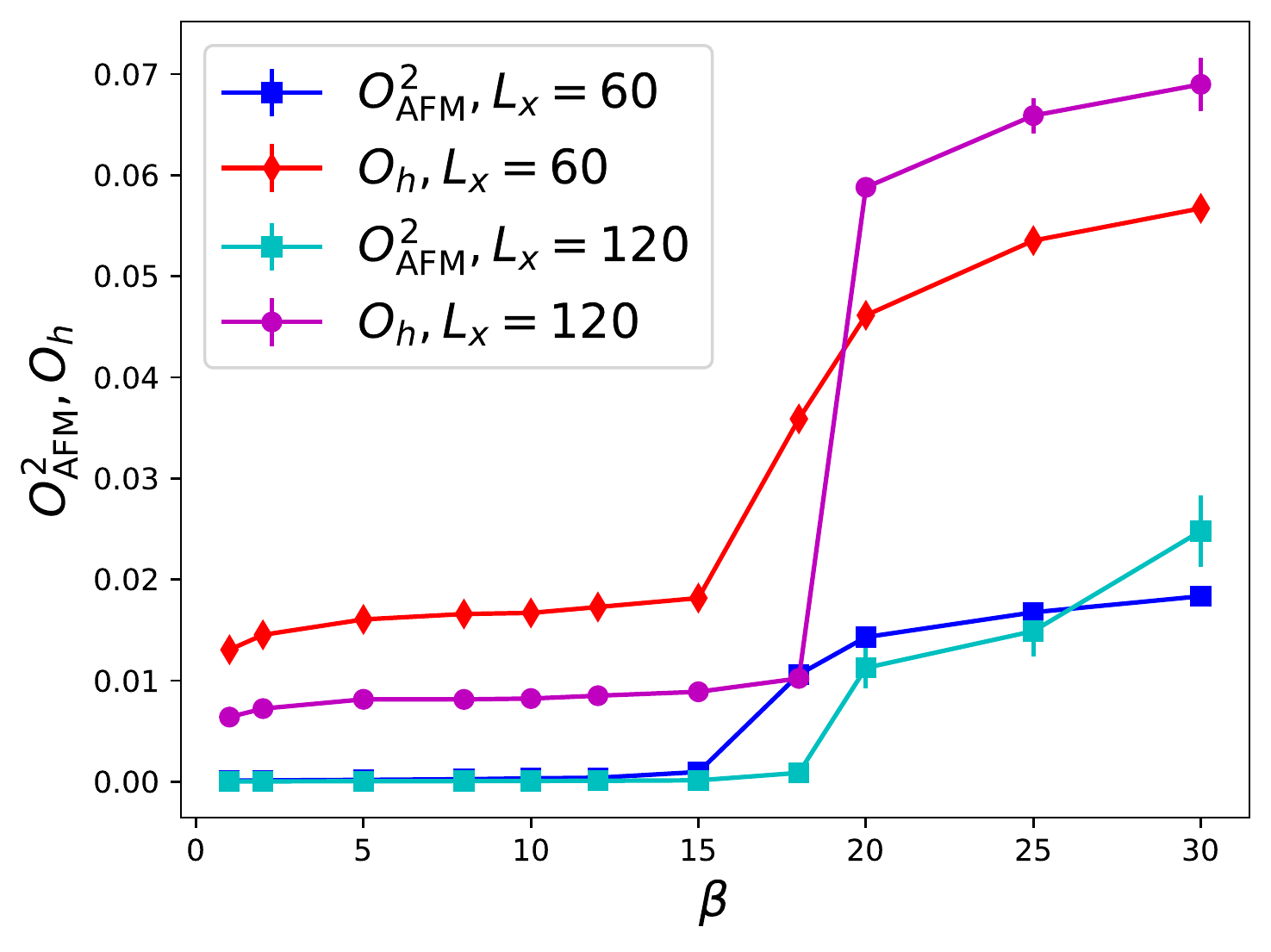}
  \caption{Antiferromagnetic  and hole ordering order parameters for the model with $\alpha=0$ and parameters $t=1, J_{\perp}=J_z=0.4,  L_y = L_x/2$, periodic boundary conditions and two different $L_x=60$ and $120$. Simulations are performed in the canonical ensemble with $33.33$\% hole doping per row ($n_h = 1/3$).
  }
  \label{fig:stripes33}
\end{figure}

At large hole doping the system orders antiferromagnetically, with a modulation along $x$ given by the hole density.
To probe for this order, we define two separate order parameters.
First, $\rm{O_{AFM}^2} = \left<  \left( \frac{1}{L_xL_y}  \sum_{\mathbf{r}} (-1)^{\phi} \hat{S}^z_{\mathbf{r}} \right)^2 \right>$, which has the meaning of the staggered magnetization squared in squeezed space~\cite{Kruis2004, Hilker2017}: the phase $\phi$ changes between $\pm 1$ each time a site is occupied by a spin, but does not change when the site is not occupied. The square is taken so that the contributions from the two possible spin ground states for a fixed hole configuration do not average out. Evaluation of the staggered magnetization squared on the full lattice instead of in squeezed space ({\it i.e.}, setting $\phi \equiv 0$) leads to a zero signal; {\it i.e.}, the magnetic order corresponds to a hidden symmetry.
Second, $ \rm{O_{h}} = \frac{1}{L_x L_y} S_{hh}(q=\frac{2\pi}{L_x} N_h)$ probes spatial order of the holes: it is proportional to the static structure factor of the connected hole-hole correlation function $\rho^{\rm conn}_{hh}(\mathbf{r}_1, \mathbf{r}_2)$, with a momentum along $x$ set by the hole density. The momentum along y is taken to be zero, implying a probe for vertical stripes. An analysis of the full momentum dependence of the static structure factor shows that this is indeed the case. We have divided the structure factor by the system volume implying that a constant value at low temperature is indicative of lattice symmetry breaking.

These two order parameters are shown in Fig.~\ref{fig:stripes20} for a hole density of $n_h = 0.2$. We see strong indications of a first order transition at $\beta t = 14(1)$, where the antiferromagnetic and the hole order parameter jump. 
The physical picture is one of vertical stripes, with a period set by the hole density along the $x$-direction. The spins are oriented in opposite directions across a hole. In the ordered phase, the simulations show broad distributions and aim to sample both spin ground states. Typically, the hole order parameter has lower error bars than the antiferromagnetic one. We also see that the transition for $L_x = 60$ is remarkably round and broad, and it becomes a lot sharper (and indicative of a single first order transition) for $L_x=120$. We see that $\rm{O_{h}}$ is strongly affected by the longer system size whereas signals in the antiferromagnetic channel appear almost system size independent. We attribute this to a reduction in hole position fluctuations as $L_y$ is increased, to which the hidden spin order is insensitive.

When we increase the hole density to  $n_h = 0.25$ and $n_h = 1/3$, as is shown in Fig.~\ref{fig:stripes25} and Fig.~\ref{fig:stripes33} respectively, we get a qualitatively similar behavior of the order parameters. The transition temperature is however slightly lower when increasing the hole density, but the first order nature of the transition more pronounced. The magnitude of the order parameters drops however rapidly between $n_h = 1/5$ and $n_h = 1/3$, by roughly a factor of 4. Note that the magnitude of the AFM order parameter for $n_h = 0.25$ and $L_x = 60$ is anomalously low in Fig.~\ref{fig:stripes25} compared to Figs.~\ref{fig:stripes20} and ~\ref{fig:stripes33}. The reason is that the superstructure requires $8$ sites for this hole doping (eg,$ \vert \uparrow, \downarrow, \uparrow, 0, \downarrow, \uparrow, \downarrow, 0 , \ldots \rangle$) and this is not commensurate for $L_x = 60$, in turn frustrating the system and leading to a reduction in magnitude of the AFM order parameter. In all other cases we considered, the superstructure is commensurate with the system size.

Compared to the mixed-dimensional $t-J$ model for fermions with $SU(2)$ spin interactions, which was recently simulated with density matrix renormalization group simulations~\cite{Schloemer2022}, our results agree on the existence of stripes with charge and hidden spin order but they also differ in a number of notable ways: (i) the presence of phase separation, attributed to the string picture and the fact that quantum fluctuations cannot mend the $J_z$ spin interactions along the $\hat{y}$-direction of the lattice; (ii) the critical temperature for stripe formation is here 3-5 times lower, and scales oppositely with the hole density; (iii) charge fluctuations were found to drive the transition in the $SU(2)$ case, while here no mechanism can be inferred. We believe that the latter two are a consequence of the fact that only the spin interactions along $\hat{y}$ favor stripe formation, whereas the quantum fluctuations along $\hat{x}$ tend to mend the N{\'e}el interactions along $\hat{x}$.

\section{Conclusion}
\label{sec:conclusion}
In conclusion, the bosonic mixed-dimensional $t-J$ models showed physics worth investigating on their own which is different from the fermionic mixed-dimensional $t-J$ or the  mixed-dimensional $t-J_z$ models.
In the model with $\alpha = 1$ the observations of an easy-plane ferromagnetic order combined with the holes moving around as free fermions, is reminiscent of spin-charge separation. 
The model with $\alpha = 0$ showed a clear sign of phase separation at low hole density -- but extending to a remarkably high $n_h = 0.14(1)$ --  and vertical stripes with antiferromagnetic easy-axis order and  a period set by the hole density for higher hole densities. The transition temperature from the disordered state into these stripes is very weakly dependent on the hole density, and the transition is first order. We could hence establish stripe order when the spin exchange term is anisotropic (nematic) for mixed-dimensional bosonic $t-J-J_z$ models, with an antiferromagnetic N{\'e}el order in squeezed space reminiscent of $1D$ systems.

{\it Acknowledgements -- } 
We wish to thank H. Schl{\"o}mer, A. Bohrdt, Z. Zhu, M. Greiner, E. Demler, and I. Bloch for fruitful discussions.
LR and LP acknowledge support from FP7/ERC Consolidator Grant QSIMCORR, No. 771891, and the Deutsche Forschungsgemeinschaft
(DFG, German Research Foundation) under Germany's Excellence Strategy -- EXC-2111 -- 390814868. FG acknowledges funding from the European Research Council (ERC) under the European Union's Horizon 2020 research and innovation program (Grant Agreement no 948141) --  ERC Starting Grant SimUcQuam.
Our simulations make use of the ALPS\-Core library~\cite{Gaenko_2017} for error evaluation. Numerical data for this paper is available under \url{https://github.com/LodePollet/QSIMCORR}.

%%%%%%%%%%%%%%%%%%%%%%%%%%%%%%%%%%%%%%%%%%%%%%%%%%%%%%%%%%
%%%%%%%%%%%%%%%%%%%%%%%%%%%%%%%%%%%%%%%%%%%%%%%%%%%%%%%%%%
%%%%%%%%%%%%%%%%%%%%%%%%%%%%%%%%%%%%%%%%%%%%%%%%%%%%%%%%%%
\bibliography{refs_mixdtj}
\end{document}